\documentclass[10pt,twocolumn,preprintnumbers,amsmath,amssymb,nofootinbib,superscriptaddress,aps,prd,numberedappendix]{openjournal}
\usepackage{newtxtext,newtxmath}
\usepackage{amsmath}
\usepackage{amssymb}
\usepackage{graphicx}
\usepackage{dcolumn}
\usepackage{bm}
\usepackage[dvipsnames]{xcolor}
\usepackage{xspace}
\usepackage{hyperref}

\newcommand{\fkem}{{\tt FKEM}\xspace}
\newcommand{\levin}{{\tt Levin}\xspace}
\newcommand{\matter}{{\tt matter}\xspace}
\newcommand{\ccl}{{\tt CCL}\xspace}
\usepackage{ulem}

\begin{document}

\preprint{APS/123-QED}

\title{The N5K Challenge: Non-Limber Integration for LSST Cosmology}

\author{
C. D. Leonard$^{1,*}$,
T. Ferreira$^{2,7}$,
X. Fang$^3$,
R. Reischke$^4$,
N. Schoeneberg$^5$,
T. Tr\"oster$^6$,
D. Alonso$^7$,
J.~E. Campagne$^8$,
F. Lanusse$^9$,
A. Slosar$^{10}$,
M. Ishak$^{11}$,
the LSST Dark Energy Science Collaboration}
\email{$^*$danielle.leonard@ncl.ac.uk}

\affiliation{$^1$School of Mathematics, Statistics and Physics, Newcastle University, Newcastle upon Tyne, NE1 7RU, United Kingdom
}
\affiliation{$^2$Laborat\'orio Interinstitucional de e-Astronomia, 20921-400, Rio de Janeiro, RJ, Brazil}
\affiliation{$^3$BCCP, Department of Physics, UC Berkeley
Campbell Hall 341, Berkeley, CA 94720, USA}
\affiliation{$^4$ German Centre for Cosmological Lensing, Astronomisches Institut, Ruhr-Universität Bochum, Universitätsstr. 150, 44801 Bochum, Germany} 
\affiliation{$^5$Institut de Ci\`encies del Cosmos, Universitat de Barcelona, Mart\'{\i} i Franqu\`es 1, Barcelona 08028, Spain}
\affiliation{$^6$Institute for Particle Physics and Astrophysics, ETH Z\"urich, Wolfgang-Pauli-Strasse 27, 8093 Z\"urich, Switzerland}
\affiliation{$^7$Department of Physics, University of Oxford, Denys Wilkinson Building, Keble Road, Oxford OX1 3RH, United Kingdom}
\affiliation{$^8$Universit\'e Paris-Saclay, CNRS/IN2P3, IJCLab, 91405 Orsay, France}
\affiliation{$^9$Universit\'e Paris-Saclay, Universit\'e Paris Cit\'e, CEA, CNRS, AIM, 91191 Gif-sur-Yvette, France}
\affiliation{$^{10}$Physics Department, Brookhaven National Laboratory, Upton, NY 11973, USA}
\affiliation{$^{11}$Department of Physics, The University of Texas at Dallas, Richardson, TX 75080, USA}
\date{\today}

\begin{abstract}
The rapidly increasing statistical power of cosmological imaging surveys requires us to reassess the regime of validity for various approximations that accelerate the calculation of relevant theoretical predictions. In this paper, we present the results of the `N5K non-Limber integration challenge', the goal of which was to quantify the performance of different approaches to calculating the angular power spectrum of galaxy number counts and cosmic shear data without invoking the so-called `Limber approximation', in the context of the Rubin Observatory Legacy Survey of Space and Time (LSST). We quantify the performance, in terms of accuracy and speed, of three non-Limber implementations: {\tt FKEM (CosmoLike)}, \levin, and \matter, themselves based on different integration schemes and approximations. We find that in the challenge's fiducial 3x2pt LSST Year 10 scenario, {\tt FKEM (CosmoLike)} produces the fastest run time within the required accuracy by a considerable margin, positioning it favourably for use in Bayesian parameter inference. This method, however, requires further development and testing to extend its use to certain analysis scenarios, particularly those involving a scale-dependent growth rate. For this and other reasons discussed herein, alternative approaches such as {\tt matter} and {\tt Levin} may be necessary for a full exploration of parameter space. We also find that the usual first-order Limber approximation is insufficiently accurate for LSST Year 10 3x2pt analysis on $\ell=200-1000$, whereas invoking the second-order Limber approximation on these scales (with a full non-Limber method at smaller $\ell$) does suffice.
\end{abstract}

\maketitle


\section{\label{sec:level1} Introduction}

In recent years, observational cosmology with photometric surveys has become a precision science, with Stage III surveys such as Dark Energy Survey \citep[DES,][]{DESDR2}, Hyper Suprime-Cam \citep[HSC, ][]{HSCDR3}, and the Kilo Degree Survey \citep[KiDS][]{KiDSDR4} measuring weak lensing and galaxy clustering at a precision which has resulted in parameter constraints rivalling the gold standard previously set by cosmic microwave background analyses \citep{Planck2018}. At the same time, the next generation of such surveys is rapidly approaching, with the Rubin Observatory Legacy Survey of Space and Time \citep[LSST,][]{Ivezic2019} poised to begin operations in 2024. Alongside its space-based contemporary {\it Euclid} \citep{Euclid2022, EuclidDef}, and the future Nancy Grace Roman Space Telescope \citep{Akeson2019}, LSST will augment galaxy samples by an order of magnitude with respect to current datasets, and correspondingly reduce statistical uncertainties on measurements of weak lensing and galaxy clustering. 

As statistical uncertainties shrink, it becomes imperative that we control sources of systematic error to a comparable degree. Otherwise, we forfeit the gains offered by the improvements in data volume of LSST, or worse, risk biases in our cosmological results. One very important set of concerns in this vein comes under the heading of accurately modelling observable quantities. 
Our purely theoretical predictions for weak lensing and galaxy clustering statistics must be more accurate than ever before; as a result, certain approximations that have historically been used no longer result in sufficient accuracy in all cases.

The key instance of this on which we focus this paper is the Limber approximation. The Limber approximation, as defined rigorously in Section~\ref{sec:theory}, is based on the assumption that the kernels of all the projected fluctuations under study vary only on scales that are much larger than the typical clustering length \citep{Limber1954}. In the context of a joint analysis which considers angular two-point auto- and cross-correlation functions (or power spectra) of weak lensing and galaxy clustering in tomographic bins (the so-called `3x2pt analysis'), the Limber approximation reduces the theoretical calculation of these quantities from a three-dimensional numerical integral over highly oscillatory Bessel functions to a one-dimensional integral over well-behaved non-oscillatory functions. This normally results in massive gains in computation speed and code simplicity with respect to brute-force approaches.

The Limber approximation is most accurate at smaller scales and fails at larger scales. It also becomes invalid for narrow kernels (thus impacting galaxy clustering statistics more strongly), especially in cross-correlations between tomographic bins with a small overlap. Historically, the uncertainties on theoretical predictions of weak lensing and galaxy clustering two-point functions as a result of making this approximation were highly subdominant to statistical uncertainties. Recently, however, this has begun to change. The DES Year 3 cosmological analysis found that non-Limber integration was required for theoretical predictions of some angular scales in their galaxy clustering data vector (the subset of 3x2pt analyses where the Limber approximation would be expected to fail first) \citep{Abbott2022}. Moving towards final analyses of Stage III surveys and into the era of LSST, we expect the use of the Limber approximation in our theoretical calculations to be less and less appropriate. It was shown, for example, in \cite{Kilbinger2017} that the standard Limber approximation achieves percent-level accuracy (a rough guideline of the modelling accuracy needed for Stage IV surveys) in the cosmic shear case only for $\ell>100$ (and we would  expect it to fail at larger $\ell$ when including galaxy clustering measurements). The suitability of the standard Limber approximation is even less obvious for these upcoming surveys if we would like to target the detection of large-scale general relativistic effects or constraints on other large-scale phenomena such as primordial non-Gaussianity.

Of course, even the most oscillatory integrand can be handled by a sufficiently aggressive brute-force numerical integration algorithm. However, in order to ultimately perform cosmological parameter inference, standard sampling algorithms require us to evaluate our theoretical predictions $O(10^{5-6})$ times. Each theoretical evaluation of our data vector must therefore be as efficient as possible; more than a couple of seconds for a single set of parameters makes the analysis extremely cumbersome and computationally expensive \citep{Torrado2021}. Implementing efficient methods for non-Limber integration is therefore of paramount importance.

Driven by the concerns and issues outlined above, there has recently been progress developing algorithms to compute these oscillatory integrals with efficiency for the specific case of weak lensing and galaxy clustering two-point functions \citep{Campagne2017,Assassi2017,Schoeneberg2018,Fang2020,Bella2021}. The goal of this paper is to determine the methods that best suit the analysis needs of LSST Dark Energy Science Collaboration (DESC) specifically, and which therefore should be incorporated in the theory prediction pipelines of DESC, based on the Core Cosmology Library, \ccl \citep{Chisari2019}. Early versions of \ccl supported the calculation of non-Limber angular power spectra for galaxy clustering only using an efficient quadrature method based on Chebyshev polynomials \citep[\texttt{Angpow},][]{Campagne2017}. However, as the complexity of the cosmological tracers for which \ccl could compute angular correlations grew, it became apparent that a more general implementation was necessary. Thus, we launched the affectionately dubbed Non-local No-Nonsense Non-Limber Numerical Knockout\footnote{`Non-local' refers to the fact that the challenge was launched during the first year of the Covid-19 pandemic, thus precluding any option of convening entrants `locally' (in person).} (N5K) Challenge\footnote{https://github.com/LSSTDESC/N5K}. The challenge was open to entrants inside and outside of the DESC and invited all participants to submit their non-Limber integration methods to be tested for accuracy and speed in the case of a specific LSST 3x2pt analysis set-up. 

In this paper, we present the results of the N5K challenge. In Section~\ref{sec:theory}, we introduce the theoretical formalism of the Limber approximation itself and the numerical integrals which must be performed. Section~\ref{sec:setup} details the specific configuration of weak lensing and galaxy clustering samples and two-point functions to be computed for the challenge, and describes the metrics by which challenge entries were to be evaluated. In Section~\ref{sec:entries}, we give an overview of each method which was entered in the challenge, and in Section~\ref{sec:results} we go on to present and discuss how each method performed in the evaluation of challenge entries. Section~\ref{sec:conc} concludes.

\section{Theory}
\label{sec:theory}
In this section, we provide the mathematical details of Limber vs non-Limber integration. To understand how this fits into the cosmological analysis set-up in question, first recall that in the case of weak lensing and photometric galaxy clustering, the key observables are the angular power spectra of cosmic shear and the projected galaxy overdensity in tomographic redshift bins. Because of the projected nature of the phenomenon of lensing, and the redshift uncertainty inherent in clustering samples without spectroscopy, we do not directly measure the 3D power spectra. Theoretical calculations of the angular power spectra of weak lensing, galaxy-galaxy lensing and galaxy clustering all require the projection along the line-of-sight direction of the 3D matter power spectrum. It is the integral which performs this projection with which we concern ourselves.

\subsection{Non-Limber power spectra}
\label{ssec:theory.nonlimber}

\noindent
We begin by describing the numerical calculation we need to tackle, before explaining how the Limber approximation is able to simplify it. A calculation of the angular power spectrum $C_\ell^{i,j}$ of two projected tracers $i$ and $j$ involves solving a triple integral of the form

\begin{align}\label{def_nonlimb_gen}
C_\ell^{i,j}=\frac{2}{\pi}\iiint &d\chi_1\, d\chi_2\, dk\, \Big(k^2\,P_{ij}(k,\chi_1,\chi_2)\, \nonumber \\ & \times \Delta_\ell^i(k,\chi_1)\Delta_\ell^j(k,\chi_2) \Big).
\end{align}
Here $\chi_1$ and $\chi_2$ are comoving distances, and $P_{ij}(k,\chi_1,\chi_2)$ is the power spectrum between the two three-dimensional fields associated with the projected tracers being correlated. The line-of-sight source functions $\Delta_\ell^i(k,\chi)$ can in general be written as
\begin{equation}\label{def_tracers_general}
  \Delta_\ell^i(k,\chi)=f_\ell^i\,K_i(\chi)\,j_\ell^{(m_i)}(k\chi),
\end{equation}
where:
\begin{itemize}
  \item $f^i_\ell$ is an $\ell$-dependent prefactor, usually associated with angular derivatives.
  \item $K_i(\chi)$ is a kernel, dependent only on redshift~$z$ or equivalently the comoving distance~$\chi$.
  \item $j_\ell^{(m_i)}(x)$ are related to the spherical Bessel functions $j_\ell(x)$ via
  \begin{equation}
    j^{(m)}_\ell(x)=\left\{
    \begin{array}{ll}
      j_\ell(x) & m=0 \\
      j'_\ell(x) & m=1 \\
      j''_\ell(x) & m=2 \\
      j_\ell(x)/x^2 & m=-2
    \end{array}\right.
  \end{equation}
\end{itemize}
The challenge of accurately integrating Eq. \ref{def_nonlimb_gen} numerically can be attributed both to the high dimensionality of the integral and to the oscillatory nature of the integrand. One can imagine a number of strategies to tackle this, many of which either invoke Fourier transforms or expansions of the integrand into basis functions \citep{Smith1974}. The power spectrum $P_{ij}$ is generally a smooth function of both $k$, $\chi_1$ and $\chi_2$. The kernels are often also smooth in $\chi$ (although this is not uniformly true for the case of galaxy clustering, particularly for the case, not considered directly here, of spectroscopic lens samples). The functions $j_\ell^{(m)}$, however, are highly oscillatory as a function of $k \chi$.

Given this, approaches to solving Eq. \ref{def_nonlimb_gen} fall into two categories. One may first calculate the integral over $k$ using fast Fourier methods, yielding a function of two variables $(\chi_1,\chi_2)$. Since the result is often non-oscillatory, the associated two-dimensional integral can be solved with simple quadrature methods. Alternatively, one may first solve the two integrals over $\chi_1$ and $\chi_2$ (e.g. invoking the approximation $P(k,\chi_1,\chi_2)\simeq\sqrt{P(k,\chi_1,\chi_1)P(k,\chi_2,\chi_2)}$ in order to make the problem factorisable). Although the result may not be non-oscillatory functions of $k$, one benefits from the fact that the final integral (over $k$) is one-dimensional, and thus potentially faster to calculate.

We focus, for the purpose of this work, on the two projected tracers of galaxy clustering (through number counts of galaxies) and weak lensing (through the shear of observed galaxy shapes). In order to simplify the definition of the calculation to be carried out as part of this numerical challenge -- without diminishing its computational complexity -- we make a number of approximations. 

First, we have discarded the effects of redshift-space distortions, magnification bias, and relativistic effects in galaxy clustering, and that of intrinsic alignments as well as the contributions from the evolution of the underlying weak lensing source galaxy density with redshift \citep[see e.g.][]{Bonvin2011}. The contribution of some of these effects, although sub-dominant, should be easily detectable by LSST. Nevertheless, the features that are primarily relevant for a non-Limber calculation of angular power spectra are already present in their respective dominant terms. Explicitly, galaxy clustering is localised in redshift, and thus its comparably narrow kernel drives the need for a non-Limber calculation. On the other hand, cosmic shear is a cumulative effect in projection, and thus $K_i$ has a much broader support in $\chi$. These two categories span the main properties of the neglected terms, and thus the challenge set-up is still representative.

Second, we also assume that the quantities involved are perfectly correlated at different cosmic times:
\begin{equation}
P_{ij}(k,\chi_1,\chi_2)=\sqrt{P_{ij}(k,\chi_1)P_{ij}(k,\chi_2)}.
\end{equation}
Although nonlinear structure formation predicts departures from this on small scales \citep{Kitching2017,Chisari2019b}, they should be strongly sub-dominant and probably undetectable for LSST. 

Third, we use simple approximations for the relation between the galaxy counts overdensity $\delta_g$, lensing potential $\Phi$, and the non-relativistic matter overdensity $\delta_m$. For galaxy clustering we assume a scale-independent linear bias relation $b_g(z)$ to the matter overdensity, such that
\begin{equation}\label{def_bias_relation}
    \delta_g(k,z) \approx b_g(z) \delta_m(k,z)~.
\end{equation}
Although this model is only accurate on large scales ($k\lesssim0.1\,{\rm Mpc}^{-1}$), more realistic models would not change the smooth nature of the scale dependence of $\delta_g$ and $\delta_m$ in both $k$ and $z$, and therefore would not invalidate the results of the challenge. For the lensing potential, we implicitly make use of a Poisson equation to relate the gravitational potential $\Phi$ to the overdensity $\delta_m$ under the assumption of negligible radiation fraction and negligible anisotropic stress:
\begin{equation}
    \nabla^2 \Phi \approx \frac{3}{2} H_0^2 \Omega_m a^{-1}\delta_m.
\end{equation}
Here $H_0$ and $\Omega_m$ are the expansion rate and the non-relativistic matter fraction today, and $a=1/(1+z)$ is the scale factor. The validity of this step breaks down when including extremely light massive neutrinos, for modifications of GR, or for non-trivial models of the cold dark matter or dark energy (all of which modify this Poisson equation). Abandoning this approximation would simply require replacing the matter power spectrum with that of the lensing potential, or introducing additional scale- and redshift-dependent factors, which again do not significantly change the smooth nature of the assumed power spectra, and thus do not invalidate the results of the challenge.

To summarise, the tracers which we consider take the following specific forms (compare to Eq. \ref{def_tracers_general}):
\begin{itemize}
  \item {\bf Galaxy number counts:}
  \begin{align}\label{kernel_g}
    f_\ell^i&=1,\hspace{12pt} m_i=0 \nonumber \\ K_i(\chi)&=H(z)\,p_i(z)\,b_g(z).
  \end{align}
  Here $H(z)$ is the expansion rate at redshift $z$ (implicitly corresponding to a comoving distance $\chi$), $p_i(z)$ is the redshift distribution of the sample, and $b_g(z)$ is the linear galaxy bias (see Eq. \ref{def_bias_relation}).
  \item {\bf Weak lensing:}
  \begin{align}\label{kernel_l}
    f_\ell^i&=\sqrt{\frac{(\ell+2)!}{(\ell-2)!}},\hspace{12pt}m_i=-2 \nonumber \\
    K_i(\chi)&=\frac{3H_0^2\Omega_m}{2a}\chi\int_z^\infty dz'\,p_i(z')\frac{\chi(z')-\chi}{\chi(z')}.
  \end{align}

Combining these two tracers either with themselves or each other leads to a 3x2pt set-up.
\end{itemize}

\subsection{The Limber approximation}\label{ssec:theory.limber}
As we can see, computing Eq.~(\ref{def_nonlimb_gen}) in practice requires the evaluation of a highly oscillatory integral over spherical Bessel functions, which -- if conducted naively through brute-force numerical integration methods -- requires considerable computational time to converge. The Limber approximation \citep{Limber1954} enables us to simplify this equation under certain circumstances. Concisely, the Limber approximation states that, because the spherical Bessel functions are highly peaked around $x\sim\ell+1/2$ and because further oscillations cause large cancellations inside of the integral \citep[for a mathematical discussion, see also][]{Loverde2008}, we can approximate
\begin{equation}
    j_\ell(x) \approx \sqrt{\frac{\pi}{2 \ell + 1}}\delta_{\rm D}\left(\ell + \frac{1}{2} - x\right)
    \label{def_limb_approx}
\end{equation}
where $\delta_{\rm D}$ is a Dirac delta function. An alternative way of stating the Limber approximation is by isolating the two Bessel functions in Eq. \ref{def_nonlimb_gen} in the inner integral (again, assuming they are narrowly peaked compared to the bandwidth of all other functions involved), and using their orthogonality relation:
\begin{equation} \label{def_limb_alt}
  \int_0^\infty dk\,k^2\,j_\ell(k\chi_1)j_\ell(k\chi_2)=\frac{2\delta(\chi_1-\chi_2)}{\pi\chi_1\chi_2}.
\end{equation}
In cases where Eq.~(\ref{def_nonlimb_gen}) contains, for a specific combination of tracers, derivatives of spherical Bessel functions, these can be written in terms of $j_\ell(x)$ using recurrence relations \citep{Goldstein1959} to invoke Eq.~\ref{def_limb_approx} or Eq.~\ref{def_limb_alt}. 

The Limber approximation of course approximates the full expression (\ref{def_nonlimb_gen}) better in some cases than in others, and is most accurate when \citep[see e.g.][]{Fang2020, Loverde2008}:
\begin{itemize}
    \item{the kernels $K_i$ and $K_j$ have a much larger extent in comoving distance $\chi$ than the physical scale which is being probed, and}
    \item{the kernels $K_i$ and $K_j$ have substantial overlap in comoving distance.}
\end{itemize}
Thus, in the case of a 3x2pt analysis, the Limber approximation breaks down in validity most readily 
\begin{itemize}
\item{when $\ell$ is small (corresponding to large angular scales),}
\item{when neither tracer $i$ nor $j$ is tracing weak lensing (because the weak lensing kernel is extremely broad in $\chi$), and}
\item{particularly for cross-correlation between different redshift bins of non-lensing tracers, where $K_i$ and $K_j$ will have less overlap.}
\end{itemize}
Note that the statement that the Limber approximation breaks down less readily for weak lensing is in reference to standard cosmic shear 2-point analysis. This would become less the case when considering alternate weak lensing analysis strategies which may alter the broad nature of the lensing kernel, such as, for example, the nulling scheme of \cite{Bernardeau2014}.

Applying the Limber approximation to Eq.~(\ref{def_nonlimb_gen}), and considering only the two tracers defined above, the integral reduces to

\begin{align}
    C_\ell^{i,j} =F^{ij}_\ell\int\frac{d\chi}{\chi^2}K_i(\chi)K_j(\chi)\,P_{ij}\left(k=\frac{\ell+1/2}{\chi},z\right),\label{limber_gen}
\end{align}
with $F^{ij}_\ell\equiv f_\ell^if_\ell^j(\ell+1/2)^{m_i+m_j}$ being an $\ell$-dependent prefactor (approximately equal to one for large $\ell$). This results in a much simpler calculation, involving a single integral and no oscillatory functions.

The goal of the N5K challenge is to compare, for accuracy and speed, methods to evaluate the full non-Limber integral (Eq. \ref{def_nonlimb_gen}) for a 3x2pt LSST-like analysis as defined in the next Section.

\subsubsection{The second-order Limber approximaton}
\label{subsubsec:secondorder}

The standard version of the Limber approximation of Eq. \ref{limber_gen} relies, as stated above, on the fact that Bessel functions $j_\ell(x)$ are highly peaked around $x=\ell+1/2$. One can therefore derive Eq. \ref{limber_gen} by Taylor expanding Eq. \ref{def_nonlimb_gen} about $\nu \equiv \ell+1/2$ and keeping only the first order term, as was made explicit in \cite{Loverde2008}. In this context, we then have the notion of a `second-order' Limber approximation, in which we keep both first and second non-zero terms in this expansion. Eq. \ref{limber_gen} becomes

\begin{align}
\label{eq:2nd_order_Limber}
    C_\ell^{i,j} &=F^{ij}_\ell\int\frac{d\chi}{\chi}\tilde{K}_i(\chi)\tilde{K}_j(\chi)\,P_{ij}\left(k=\frac{\nu}{\chi},z\right)  \Bigg\{1 \nonumber \\ &+ \frac{1}{\nu^2}\Bigg[\frac{\chi^2}{2}\Bigg(\frac{\tilde{K}_i''(\chi)}{\tilde{K}_i(\chi)} + \frac{\tilde{K}_j''(\chi)}{\tilde{K}_j(\chi)} \Bigg) \\ & \quad\;\;\; + \frac{\chi^3}{6}\Bigg(\frac{\tilde{K}_i'''(\chi)}{\tilde{K}_i(\chi)} + \frac{\tilde{K}_j''(\chi)}{\tilde{K}_j(\chi)} \Bigg) \Bigg] \Bigg\},\nonumber\label{limber_second}
\end{align}
where $\tilde{K}_i(\chi) \equiv K_i(\chi) / \sqrt{\chi}$ and primes indicate derivatives with respect to $\chi$ (which would typically be computed numerically when implementing this expression). The equivalent expression in \cite{Loverde2008} assumes a spin-0 field (e.g. clustering) and thus sets $F_\ell^{ij}=1$; \cite{Kilbinger2017} clarify this point for the spin-2 case of cosmic shear.
Lastly, one should note that the series expansion in Eq. \ref{eq:2nd_order_Limber}, is not a convergent series over all $\ell$ as noted in the original paper \citep{Loverde2008}. The range of validity depends on the position and width of the kernels $\tilde{K}(\chi)$. That being said, using higher order terms over a divergent range in multipoles can and will lead to worse results.

\section{Challenge set-up}
\label{sec:setup}

\subsection{Analysis configuration}
\label{subsec:config}

\begin{figure*}
  \centering
  \includegraphics[width=0.9\textwidth]{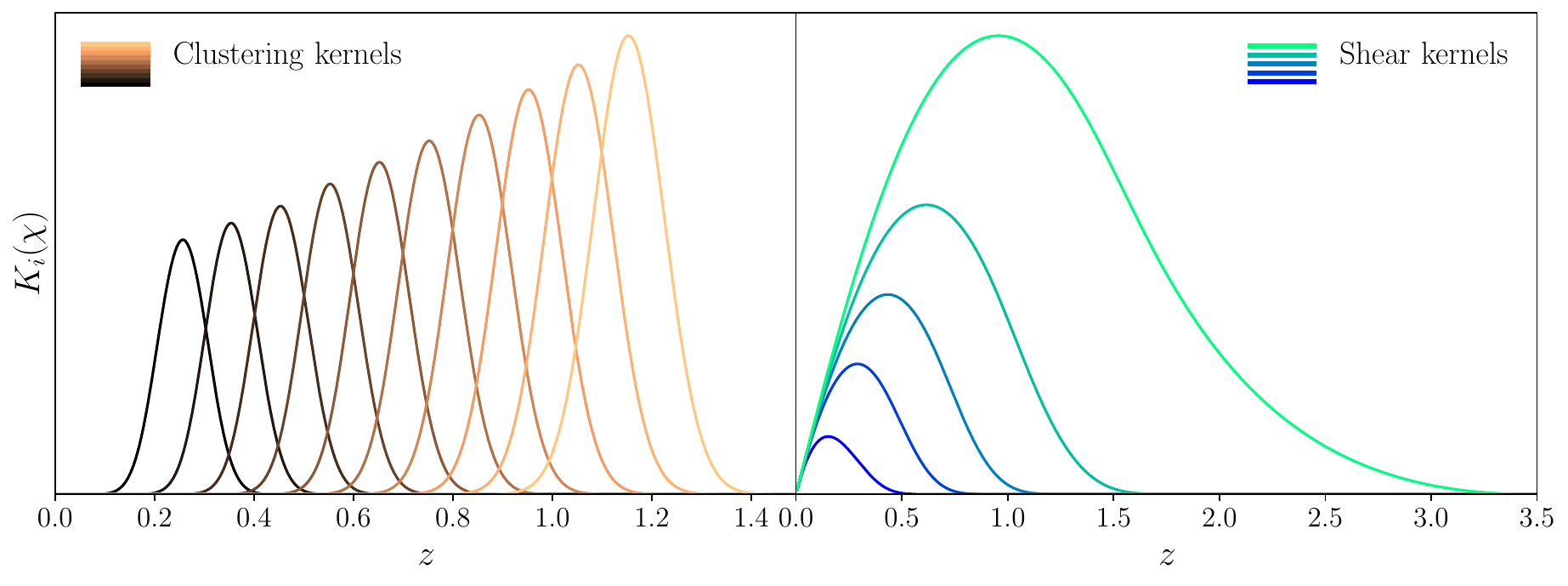}
  \caption{Fiducial kernels associated with the 10 galaxy clustering bins (left) and the 5 cosmic shear bins (right) used in this work, corresponding to the LSST-DESC Y10 scenario used in \citep{DESCSRD}.}
\label{fig:kernels}
\end{figure*}
The fiducial analysis configuration for the challenge is a so-called 3x2pt analysis. A 3x2pt analysis jointly incorporates the two-point functions of galaxy clustering, galaxy-galaxy lensing, and cosmic shear, all in multiple tomographic redshift bins. In Fourier space, these two-point functions are given by Eq.~(\ref{def_nonlimb_gen}) in the full non-Limber case.

We select a specific analysis configuration which mimics a simplified version of the LSST Year 10 analysis scenario in the LSST DESC Science Requirements Document \citep[SRD,][]{DESCSRD}. Specifically:
\begin{itemize}
    \item{We assume a fiducial $\Lambda$CDM cosmology with parameter values of $\Omega_{\rm m}=0.3156$, \mbox{$\Omega_{\rm b}=0.0492$}, \mbox{$w_0 = -1.0$}, $H_0=67.27$ km/s/Mpc, \mbox{$A_s = 2.12107 \times 10^{-9}$}, and $n_s = 0.9645$.
    }
    \item{We consider 10 tomographic redshift bins of the clustering sample and 5 tomographic redshift bins of the shear sample. These are defined according to the specifications given in Appendices D1 and D2 of \citep{DESCSRD}, including photo-z uncertainty. Briefly, we impose a top-hat binning on the underlying assumed true redshift distribution, and then convolve with a Gaussian photo-z uncertainty model with $\sigma_z=0.03(1+z)$ for lenses and $\sigma_z=0.05(1+z)$ for sources. The kernels associated with these redshift bins are shown in the left and right panels of Fig. \ref{fig:kernels} for clustering and shear respectively. The number density and shape noise of these samples, which determines the noise component of each tracer, also follow the DESC SRD.}
    \item{We consider all possible auto- and cross- spectra. This results in a total of 15 unique cosmic shear spectra (5 auto, 10 cross), 50 unique galaxy-galaxy lensing spectra (all cross), and 55 unique galaxy clustering spectra (10 auto and 45 cross), for a total of 120 unique spectra. It is worth noting that clustering cross-correlations have commonly been discarded in 3x2pt analyses, since they contain less cosmological information than the auto-correlations and are more prone to biases from photometric redshift uncertainties. Our analysis, which penalises methods that mispredict the clustering cross-correlations, is therefore conservative in this regard.}
    \item{We initially evaluated each spectrum at 103 values of $\ell$ (bandpowers), defined by imposing logarithmic spacing between $\ell=2-2000$, casting the resulting $\ell$ values as integers and retaining only unique integer values. However, during the challenge evaluation process it became clear that the more efficient and appropriate regime in which to compare entries from an accuracy perspective was solely at $\ell \le 200$. Further detail on this choice is given in Section~\ref{subsec:metrics}.} 
    \item{As described in the SRD, we assume an evolving linear bias of the form $b(z)=0.95/D(z)$, where $D(z)$ is the scale-independent linear growth factor. This has been shown to be a good approximation for magnitude-limited samples \citep{Nicola2020}.}
    \item{As mentioned in the previous section, we do not include the effects of intrinsic alignment, redshift-space distortions, or magnification. We expect that including these effects would not significantly alter the timing or accuracy of non-Limber integration.}
\end{itemize}

Challenge participants were provided with the precomputed kernels (i.e. $K_i(\chi)$ in Eqs.~(\ref{kernel_g}), (\ref{kernel_l}) above), as well as the matter power spectra and linear bias required to compute the auto- and cross-spectra detailed above. These quantities were computed using CCL. Participants were also provided with redshift distributions corresponding to the fiducial scenario should they prefer to use these as input rather than kernels.

\subsection{Benchmarks}
\label{subsec:benchmark}
In order to quantify the accuracy of the different methods, a baseline `truth' was required. A set of `benchmark' calculations were performed for the fiducial scenario to act as the true values of the computed non-Limber integrals, and the results of these calculations were provided to participants prior to the deadline of the challenge in order to check and improve the accuracy of their entries.

These benchmarks were obtained using a simple brute-force integration scheme in three dimensions. Fixing $k$ in Eq. \ref{def_nonlimb_gen}, we first integrate over $d\chi_1$ and $d\chi_2$ independently, using a standard Riemann integral with a linear interval in $\chi$ given by
\begin{equation}
  \delta\chi={\rm Min}\left(\alpha\frac{\pi}{k},\delta\chi_0\right),
\end{equation}
with $\alpha=0.05$ and $\delta\chi_0=5\,{\rm Mpc}$. Then, the integrand over $k$ was interpolated in $\log k$ using an Akima spline evaluated at $N_{\log k}=200$ samples per $e$-fold in the $\ell$-dependent interval
\begin{equation}
  \frac{\ell+1/2}{\chi_{\rm max}}<k<\frac{\ell+1/2}{\chi_{\rm min}},
\end{equation}
with $\chi_{\rm min}=13\,{\rm Mpc}$ and $\chi_{\rm max}=7000\,{\rm Mpc}$. The final integral over $\log k$ was carried out using spline integration.

We validated the benchmarks generated with this method in two different ways. First, we studied the stability of the benchmarks with respect to changes in $\alpha$, $\delta\chi_0$, $N_{\log k}$, $\chi_{\rm min}$, and $\chi_{\rm max}$, doubling or halving these parameters. We also studied the impact of using the {\tt QAG} (adaptive Gaussian quadrature) scheme, originally implemented in {\tt QUADPACK} \citep{Quadpackbook}, instead of spline integration when performing the integral over $k$. The results were shown to be remarkably stable (within $\sim0.1\%$) with respect to these changes. Secondly, we compared the benchmarks obtained for some of the galaxy-galaxy cross-correlations with the prediction from {\tt Angpow} \citep{Campagne2017}, a fast non-Limber method that was not included in the challenge\footnote{{\tt Angpow} was not included in the challenge due to its current lack of functionality for integration involving weak lensing kernels, and a lack of available development effort in porting the native C++ codebase to a python or python-wrapped version, as required for the challenge.}. The benchmarks were found to be in good agreement with {\tt Angpow}. It is worth noting that, with these very high-accuracy settings, and using the brute-force method we just described, it took approximately $60$ hours to compute all power spectra using 12 Intel cores in parallel. This illustrates the complexity of the problem, and the need to develop more efficient methods.

The accuracy of each method was quantified in terms of the $\chi^2$ statistic with respect to the benchmarks:
\begin{equation}
  \Delta\chi^2\equiv\sum_b N_b{\rm Tr}\left[\left(\Delta {\sf C}_b\,\bar{\sf C}^{-1}_b\right)^2\right],
  \label{deltachi2}
\end{equation}
where $\bar{\sf C}_b$ is a $15\times15$ matrix containing all benchmark auto- and cross-power spectra in bandpower $b$ and adding where appropriate terms accounting for LSST-level shape-noise ($\epsilon_s^2 / n_s$ with $\epsilon_s=0.28$ and $n_s$= 27 source galaxies per arcmin$^2$) and shot-noise ($ 1 / n_l$ with $n_l$=40 lens galaxies per arcmin$^2$). $\Delta{\sf C}_b$ is the difference between the benchmark prediction and the prediction from a given non-Limber method, and $N_b$ is the effective number of modes for that bandpower:
\begin{equation}
  N_b=\frac{f_{\rm sky}}{2}\sum_{\ell\in b}(2\ell+1),
\end{equation}
with $f_{\rm sky}=0.4$ the sky fraction for LSST. Note that this is equivalent to the standard definition of $\chi^2=\Delta{\bf x}^T{\sf Cov}^{-1}\Delta{\bf x}$ where $\Delta{\bf x}$ is a vector containing the difference between the predicted power spectra and the benchmarks, and ${\sf Cov}$ is the \emph{Gaussian} covariance of these power spectra \citep{Hamimeche2008}. Although we do expect non-Gaussian features in the 3x2pt covariance to be relevant for LSST, the choice to use a Gaussian covariance here is both expedient (the Gaussian covariance being of course far more straightforward to calculate, as demonstrated in Eq. \ref{deltachi2}) and conservative: by neglecting non-Gaussian terms in the covariance we are requiring entries to conform more closely to the benchmarks, rather than less.

\subsection{Evaluation metrics}
\label{subsec:metrics}
Theoretical predictions for a cosmological analysis have two main requirements: speed and accuracy. Both are crucial given the high dimensional parameter space over which cosmological inference must be performed (and hence the number of times theoretical predictions must be made), and the high statistical power of current and upcoming late-time cosmological data. The N5K challenge therefore seeks to jointly optimise these metrics. How to balance these two considerations is not entirely straightforward: if the time required for non-Limber integration is significantly below that of another step in the theory pipeline (e.g. estimating the 3D power spectra $P(k,z)$), there is minimal gain in increasing the speed further. Similarly, if the accuracy of the prediction is such that it is not expected to produce a bias in cosmological parameter inference, there is little need to further improve the accuracy. In most numerical methods, one can trade speed for accuracy, so it was essential to fix some common method for evaluation.

For the fiducial challenge evaluation, we took the approach of setting a fixed, quite strict accuracy requirement which would easily meet our analysis accuracy needs, and then comparing run time of entries at this accuracy. In the event that none of the entries could achieve run times comparable to other steps in the theory pipeline at this accuracy setting, we would revisit the accuracy requirement and adjust to a less stringent option which still met our needs. Given the experimental sensitivity of LSST Y10, we initially set the requirement to be $\Delta \chi^2 \leq 1$ over a range of $2 \leq \ell \leq 2000$. This requirement was selected on the basis that a shift in $\chi^2$ of 1 is the minimum value which could in principle correspond to an erroneous $1\sigma$ detection of some new physics which is in reality not present. This is the scenario we are seeking to avoid by ensuring we have sufficiently accurate theoretical modelling. Determining the precise level of cosmological parameter bias that a given specific implementation of non-Limber integration imposes could be achieved by a full simulated analysis taking the benchmarks simulated data vectors as truth; we do not attempt this here, relying instead on the conservative argument above, but expect such a check to be valuable in pipeline validation of future analyses.

Initially, at the close of the official challenge entry period, one entry ({\tt Levin}, see Section~\ref{sec:entries}) achieved the accuracy requirement. Investigating the difference between the various entries, it was soon found that {\tt Levin} used the extended second-order Limber approximation \citep{Loverde2008} above $\ell=200$ (implementing the required derivatives of the matter power spectrum via splines). The other two entries relied on the in-built CCL Limber approximation above $\ell=200$, where conventionally the standard first-order Limber approximation might be expected to be highly accurate (e.g. \cite{Kilbinger2017} found percent-level accuracy at $\ell>100$ for cosmic shear 2-point functions). It was this standard Limber approximation which caused a $\Delta \chi^2>1$ at $\ell\geq 200$ alone. It was determined that due to the very tight statistical uncertainty expected for LSST Y10 on redshift-bin autocorrelation spectra for galaxy clustering at $200 \leq \ell \leq 1000$, the order of the Limber approximation indeed had a great impact. Since any of the entries could in principle use the extended Limber approximation at $200 \leq \ell \leq 1000$, it was decided that the target accuracy should be $\Delta \chi^2 \leq 0.2$ at $\ell < 200$. This corresponds to the original requirement of $\Delta \chi^2 \leq 1$ over  $\ell \leq 2000$ when the extended Limber approximation is used for $200 \leq \ell \leq 1000$ and the standard Limber approximation is used on $1000 < \ell \leq 2000$.

As detailed in Section~\ref{sec:results}, after the initial evaluation with this metric, we also explored, in collaboration with participants, other secondary metrics including how each entry scaled with accuracy requirement,  number of spectra to calculate, and width of redshift bins. We detail these secondary metrics in Section~\ref{sec:results} below.

All challenge evaluations were, unless otherwise specified, run on a single Haswell compute node of the National Energy Research Scientific Computing Center (NERSC) Cori. Cori is a Cray XC40 and Haswell is an Intel Xeon Processor E5-2698 v3. By default, entries could make use of multi-threading with up to 64 threads in this node, although we will report on the scaling of the different methods with thread count.

\section{Challenge entries}
\label{sec:entries}
The N5K challenge opened on October 29, 2020 with a deadline of January 15, 2021, meaning the challenge was initially open for 79 days. The challenge was open to all, including non-DESC members. It was advertised primarily within DESC, however the challenge repository was publicly visible on {\tt github}, and challenge organisers reached out to invite entries from researchers who were known to have authored public code to implement non-Limber integration or to solve related problems; the final entrants were a mixture of DESC and non-DESC members. In January 2021, an issue was raised as to a possible bug in the benchmarks; the deadline was thus extended to February 5, 2021 in order to allow time to investigate and resolve this. 
At the final challenge deadline, seven entries had been received. Upon beginning the evaluation phase, it became clear that three of these entries were incomplete, i.e. they did not in their submitted form perform the challenge task as stated (to any accuracy requirement). One of these entries was indicated to have been started as a `placeholder submission' with intention to be updated later and contained very little code. The other two were in-progress but incomplete attempts to significantly modify existing code bases to perform similar but not the same calculations. The challenge organisers reached out to these entrants to determine whether they would be able and willing to update their entries to a complete status on a short timescale, however in all cases this invitation was declined citing lack of time to do so. This left four complete entries. Subsequently, one of the remaining four entries was withdrawn as during the analysis phase (which was performed openly with ongoing participation and communication with entrants), it became clear that this entry would require considerable overhaul to be competitive in run time or accuracy with the others. Thus, the final challenge slate comprised three entries.

In this section, each of these three algorithms is described. It is worth emphasizing here that while we discuss the mathematical algorithms associated with each entry, our evaluation of each entry is intimately tied to the specific implementation of the algorithm. The N5K challenge is thus both an algorithm challenge and one of efficient coding.

\subsection{Challenge entry: {\tt matter}}
\label{subsec:class}
\noindent
{\it Entrant: author NS}

The idea of using the FFTlog algorithm \citep{Talman78} for cosmological angular statistics was first proposed in \cite{Assassi2017}. This method is based on approximating a function $f(x)$ through complex power laws as 
\begin{equation}
    f(x) = \sum_{n=0}^N c_n x^{\nu_n}
    \label{eq:pl_mat}
\end{equation}
where the $\nu_n$ are the complex power law exponents. Usually, these are chosen as $\nu_n = 2\pi i \cdot n/N + \eta$, where $\eta$ is an arbitrary real number\footnote{The precise choice of $\eta$ can be found in \cite{Schoeneberg2018} for \texttt{matter} and \cite{Fang2020} for \fkem, and is related to aspects of the rapidity of convergence of the underlying integrals.} and $N$ is an arbitrary integer. By expressing the power law as $x^{\nu_n} = \exp(2\pi i \cdot n/N \cdot \log x) \cdot x^\eta$,
we can observe that this particular choice directly corresponds to the Fourier transform in $\log x$ and allows us to make use of the computational advantages inherent to the Fast-Fourier-Transform (FFT) algorithm. Additionally, the integrals over products of Bessel functions and power laws can usually be integrated analytically, resulting in Gamma functions (for a single Bessel function, described in Section~\ref{subsec:fkem}) or hypergeometric functions (for two Bessel functions, described in this Section~\ref{subsec:class}). As such, in either case, no strongly oscillating integrals have to be computed.

The initial implementation of the FFTlog algorithm of \cite{Assassi2017} was performed in \cite{Schoeneberg2018}. Here the power spectrum is itself expanded $P_{ij}(k,\chi_1,\chi_2) = \sum c^{i,j}_n(\chi_1,\chi_2) k^{\nu_n}$ and the corresponding power laws in $k$ times the product of two Bessel functions (see Eq.~(\ref{def_nonlimb_gen})) are integrated analytically, leading to integrals of the form \mbox{$I_\ell(\nu,t) = 4 \pi \int u^{\nu-1} j_\ell(u) j_\ell(u t) \mathrm{d}u$}, which can be expressed through the hypergeometric functions. See \citep{Assassi2017,Schoeneberg2018} for more details on the precise implementation. Importantly, since these hypergeometric functions are independent of cosmology, they can be pre-computed and efficiently tabulated. In the end, using this method only two integrals (over $\chi_1$ and $\chi_2$, or $\chi_1$ and $t=\chi_2/\chi_1$) have to be computed, involving no slowly decaying oscillatory functions. 

Explicitly, the angular power spectrum can be written as (see \citep{Schoeneberg2018})
\begin{equation}
    C_\ell^{i,j} = \sum_{n=0}^N \int_0^1 \mathrm{d}t I_\ell(\nu_n, t) \left[ f^{i,j}_n(t) + t^{\nu_n-2} f^{i,j}_n(1/t) \right]
\end{equation}
with $f^{i,j}_n(t) = \int_0^\infty \mathrm{d}\chi K_i(\chi) K_j(\chi t) c_n^{i,j}(\chi,\chi t) \chi^{1-\nu_n}$ and the coefficients $c_n(u,v)$ are computed as the Fourier expansion $g(k, u, v) = \sum k^{\nu_n} c_n(u,v)$ through the FFT algorithm in $\log k$.
As long as $N\gg 1$ and all remaining variables of integration (such as $\chi$\,, $t$) are sampled sufficiently densely, this algorithm will always accurately produce the numerical result. In practice, even choices of $N \approx 100$ are sufficient for precise results. Note that although the specification of the N5K challenge assumes a scale-independent galaxy bias, the {\tt matter} method could in principle naturally account for any relevant effects from scale-dependent galaxy bias by performing the power law expansion on the 3D galaxy power spectrum directly (and correspondingly removing the linear bias term from $K_i(\chi)$.)

\subsection{Challenge entry: {\tt FKEM (CosmoLike)}}
\label{subsec:fkem}
\noindent
{\it Entrant: author XF}

The FKEM algorithm, proposed in \citep{Fang2020} and first implemented and tested within the CosmoLike framework \citep{Krause2017}, expands upon the idea of Section~\ref{subsec:class} by re-ordering the integrals of Eq.~(\ref{def_nonlimb_gen}), separating the $\chi_1$ and $\chi_2$ integrals into two single-Bessel transforms which can be efficiently computed with the FFTlog algorithm and its extensions developed in \citep{Fang2020}. The implementation adopts a complex power-law decomposition form similar to that in Section~\ref{subsec:class}, which was first introduced in \citep{fastpt,fastpt2}.

In order to make the separation, the method splits the matter power spectrum into a linear separable component $P_{\rm lin}$ and a non-separable component

\begin{align}
P_\delta(k,z(\chi_1),z(\chi_2) &= D(\chi_1) D(\chi_2) P_{\rm lin}(k,z=0) \nonumber \\ &+ [P_{\delta}-P_{\rm lin}](k,\chi_1,\chi_2),
\label{eq:fkem_factor}
\end{align}
where $D(\chi)$ is the linear growth factor (assumed here to be scale-independent - an assumption which holds to a good approximation in $\Lambda$CDM but which breaks down at some level in the presence of massive neutrinos and in some alternative cosmological models e.g. some alternative theories of gravity), and we refer to $[P_{\delta}-P_{\rm lin}]$ as the {\it nonlinear residual}. The integral containing $P_{\rm lin}$ only can be turned into single-Bessel transforms, while that containing the nonlinear residual can be approximately modelled using the Limber approximation, introducing only sub-percent errors on large scales \citep[see][]{Fang2020} for which cosmic variance is large in any case. As shown in Section \ref{sec:results}, for the considered experimental specifications we can explicitly demonstrate this point.

Explicitly, the power spectra of tracers $i$ and $j$ are computed as

\begin{align}
    C_\ell^{i,j(ab)} &= f(\ell)\int_0^\infty \frac{\mathrm{d}k}{k} k^3 P_\mathrm{lin}(k,z=0) I_{i,\ell}^a(k) I_{j,\ell}^b(k)\nonumber\\
    &+\frac{\pi}{2}\frac{f(\ell)}{(\ell+\frac{1}{2})^{2x}} \times\nonumber\\
    &\quad\int_0^\infty \mathrm{d}\chi \frac{K_i(\chi)K_j(\chi)}{\chi^2}[P_\delta-P_{\rm lin}]\left(\frac{\ell+1/2}{\chi},z(\chi)\right),
\end{align}
where the second term is just the Limber Eq.~(\ref{limber_gen}) replacing the full nonlinear matter power spectrum with the nonlinear residual. The tracer-dependent $I_{i,\ell}^a(k)$ integrals are given by
\begin{equation}
    I_{i,\ell}^a(k) = \int_0^\infty \frac{\mathrm{d}\chi}{\chi}\frac{\chi K_i(\chi)D(\chi)}{(k\chi)^{2X_i}}j_\ell(k\chi)~,
\label{eq:I_fkem}
\end{equation}
where $X_i=0$ when $i$ corresponds to galaxy number counts, and $X_i=1$ when $i$ corresponds to weak lensing shear. We sample $\chi$ logarithmically (assuming we can effectively take the limits of the integral as $0<\chi_{\rm min} <\chi_{\rm max}<\infty$) and denote its $n$-th element as $\chi_n$ ($n=0,1,\cdots,N-1$). Assuming power-law decomposition and adopting the notation of Section~\ref{subsec:class},
\begin{equation}
    \chi_n K_i(\chi_n)D(\chi_n) = \frac{1}{N}\sum_{m=0}^{N}c_m \chi_0^\eta \left(\frac{\chi_n}{\chi_0}\right)^{\nu_m}~,
\end{equation}
equation~(\ref{eq:I_fkem}) is solved as
\begin{equation}
    I_{i,\ell}^a(k_n) = \frac{\sqrt{\pi}}{4k_n^\eta}{\rm IFFT}[\lbrace c_m^*(k_0\chi_0)^{\nu_m-\eta}g^{X_i}_\ell(\nu_m^*)\rbrace]_n~,
\end{equation}
where IFFT stands for the inverse fast Fourier transform, $k_n$ is the $n$-th element of the logarithmically sampled $k$ array with the same logarithmic spacing as the $\chi$ array (i.e. $\Delta (\ln k)=\Delta (\ln \chi)$), and the function $g^{X_i}_\ell(z)$ is given by
\begin{equation}
    g^{X_i}_\ell(z) =
      2^z\frac{\Gamma[(\ell+z)/2]}{\Gamma[(3+\ell-z)/2]}\frac{1}{[(\ell-2+z)(3+\ell-z)]^{X_i}}.
\end{equation}

The FKEM algorithm thus takes two FFTs to compute each of the $I_{i,\ell}^a(k)$ integrals for an array of $k$, and a Riemann sum over $k$ for the final $k$ integral. Note that for the $X_i=1$ case, we do not move the $k$ dependence in the denominator of equation~(\ref{eq:I_fkem}) out of the $\chi$ integral; instead, we modify the FFTlog method by introducing an additional factor in the $g_\ell(z)$ function. This greatly improves the numerical stability and is detailed on the GitHub webpage of FFTLog-and-beyond\footnote{\url{https://github.com/xfangcosmo/FFTLog-and-beyond/blob/master/Notes.pdf}}.

The algorithm can be extended in the following ways to deal with more complex scenarios:\\
\begin{enumerate}
\item{For nonlinear galaxy bias models, we may encapsulate the extra nonlinear terms into $P_\delta$, assuming that the nonlinear contribution stays in small scales where the Limber approximation is sufficiently accurate. This approach has been validated and used in the recent DES Y3 analysis \citep[see Section III B of][for detailed implementation]{Krause2021}, and can be applied to nonlinear modelling of weak lensing observables as well.}
\item{For scale-dependent linear power spectra, we may split each redshift bin into a few narrower bins, within which the growth factor is approximately scale independent. Section 2.2.2 of \citep{Fang2020} discusses this approach in depth.}
\end{enumerate}
\subsection{Challenge entry: {\tt Levin}}
\label{subsec:levin}
\noindent
{\it Entrant: authors RR and TT}

Levin's method \citep{Levin1996, Levin1997} effectively casts the quadrature problem of the oscillatory integral into the solution of a system of ordinary and linear differential equations. By choosing an appropriate basis for the solution (e.g. polynomials) the solution is constructed at points in the integral's domain, called {\it collocation} points, as a solution of a simple linear algebra problem. 

From Eq.~(\ref{def_nonlimb_gen}) one finds that the oscillatory integrals are of the form

\begin{align}\label{eq:i1}
I [h(\chi,k)] = \int_{\chi_1}^{\chi_2} \mathrm{d} \chi \,  h \left( \chi, k \right) j_\ell \left( k \chi \right)\;,
\end{align}
so that $I$ is a functional of $h(\chi,k)=K_i(\chi) \sqrt{P_{ij}(k,z(\chi))} (k\chi)^{m_i}$, which in turn can depend on any external arguments not integrated over. Note that this method, like {\tt matter}, naturally allows for the possibility of scale-dependent galaxy bias as the full 3D galaxy power spectrum can be straightforwardly incorporated in $h(\chi,k)$.
In general Levin's method is applicable to any integral of the form:

\begin{align}\label{scalarprod}
I[{\bf F}] = \int_a^b \mathrm{d} x \left\langle {\bf F}, {\bf w} \right\rangle (x)\ \equiv \int_a^b \mathrm{d} x \, {\bf F}^T (x) {\bf w}(x);	,
\end{align}

\begin{align}\label{derivata}
{\bf w}'(x) = {\sf A}(x)\,{\bf w}(x)\;.
\end{align}
It is key that the components of ${\sf A}(x)$ must not be highly oscillatory. The integrand in Eq.~(\ref{scalarprod}) is approximated by constructing a vector ${\bf p}$ satisfying

\begin{align}
\left\langle {\bf p}, {\bf w} \right\rangle' = \left\langle {\bf p}' + {\sf A}^T{\bf p}, {\bf w} \right\rangle \approx \left\langle {\bf F}, {\bf w} \right\rangle\;.
\end{align}
by applying the product rule for derivatives and exploiting Eq.~(\ref{derivata}), with $\left\langle {\bf p}, {\sf A} {\bf w} \right\rangle = \left\langle {\sf A}^T {\bf p}, {\bf w} \right\rangle$. If the vector ${\bf p}$ can be found, then the integral becomes trivial and Eq.~(\ref{scalarprod}) can be approximated by

\begin{align}
I[{\bf F}] \approx \int_a^b \mathrm{d}x \left\langle {\bf p}, {\bf w} \right\rangle ' (x) = \left\langle {\bf p}, {\bf w} \right\rangle (b) - \left\langle {\bf p}, {\bf w} \right\rangle (a)\;.
\end{align}
In practice, this translates to demanding that
$\left\langle {\bf p}, {\bf w} \right\rangle' = \left\langle {\bf F}, {\bf w}\right\rangle$, at $n$ collocation points $x_j , j = 1, 2, . . . , n$, yielding

\begin{align}
  \left\langle {\bf p}' + {\sf A}^T {\bf p} - {\bf F}, {\bf w} \right\rangle (x_j) = 0, \quad j = 1, ..., n
\end{align}
the vector $({\bf p}' + {\sf A}^T {\bf p} - {\bf F})$ thus must be orthogonal to ${\bf w}$ at the points $x_j$. The trivial solution is the null vector:

\begin{align}\label{zeros}
{\bf p}'(x_j) + {\sf A}^T (x_j) {\bf p}(x_j) = {\bf F}(x_j).
\end{align}
The vector ${\bf p}$ can be found by choosing $n$ differentiable basis functions $u_m(x)$ that ${\bf p}$ can be expanded in:

\begin{align}
p_i(x) = c_i^{(m)} u_m(x), \quad i= 1,...,d; \quad m = 1,...,n.
\end{align}
Here and in the following we sum over repeated indices.
Plugging the expansion into Eq.~(\ref{zeros}) implies a linear system of equations for the $d \times n$ coefficients ${\bf c}_i^{(m)}$:

\begin{align}\label{740}
  c_i^{(m)}u_m'(x_j) + A_{q i} c_q^{(m)} u_m(x_j) = F_i(x_j),
\end{align}
where $i,q = 1, ...,d; j,m = 1, ...,n$. 
Accuracy below $10^{-6}$, which is roughly the accuracy needed for the  challenge presented here, can often be achieved with less than 10 such {\it collocation} points.
Here we use equidistant collocation points $x_j$
and use the $n$ lowest-order polynomials as basis functions:

\begin{align}
u_m(x) = \left( \frac{x- \frac{a+b}{2}}{b-a} \right)^{m-1}, \quad m=1, ...,n\;.
\end{align}
The vector ${\bf w}$ for the integral in Eq.~(\ref{eq:i1}) can be identified by considering the recurrence relations for the spherical Bessel functions

\begin{align}
\frac{\mathrm{d}}{\mathrm{d}x} j_\ell(x) &= j_{\ell-1}(x) - \frac{\ell+1}{x} j_\ell (x)\;, \\
\frac{\mathrm{d}}{\mathrm{d}x} j_{\ell-1}(x)&= -j_\ell (x) + \frac{\ell-1}{x} j_{\ell-1}(x)\;.
\end{align}
Rewriting these relations in the form of Eq. \ref{derivata}, with $\chi$ playing the role of $x$, we find

\begin{align}
 {\bf w}(\chi) = \left( \begin{matrix}
j_\ell(k \chi) \\
j_{\ell-1}(k \chi)
\end{matrix} \right), \quad {\sf A}(\chi) = \left( \begin{matrix}
- \frac{\ell+1}{\chi} & k \\
-k & \frac{\ell-1}{\chi}
\end{matrix} \right)	 
\end{align}
is a suitable choice for the integral in Eq.~(\ref{eq:i1}), with ${\bf F}(\chi) = [h(k, \chi), 0]$. In practice the $\chi$-integration is always carried out over the support of the provided kernels $K_i(\chi)$. This is the starting interval which is then divided into $n$ collocation points. The interval is then bisected and each sub-interval is divided into $n$ collocation points again and the integral is approximated by the solution to Eq. (\ref{740}). Repeating this procedure on the interval with the largest error until convergence is reached yields the final result. The remaining $k$-integration is carried out with an adaptive integration scheme provided by the Gnu Scientific Library\footnote{https://www.gnu.org/software/gsl/} ({\tt{GSL}}).

The implementation of this method evaluated in the N5K challenge is available on github\footnote{https://github.com/rreischke/nonLimber}.

\section{Results and Discussion}
\label{sec:results}
\noindent
We first present the performance of each entry with respect to the fiducial evaluation metric as described above in Section~\ref{subsec:metrics}. We then explore how the performance of each method and implementation scales as we vary the analysis setup along several axes: number of shear and clustering bins in the 3x2pt analysis, width of bins in redshift, accuracy requirement, and  number of computing cores available.

\subsection{Fiducial results}
\label{subsec:fidres}
\begin{figure*}
  \centering
  \includegraphics[width=0.85\textwidth]{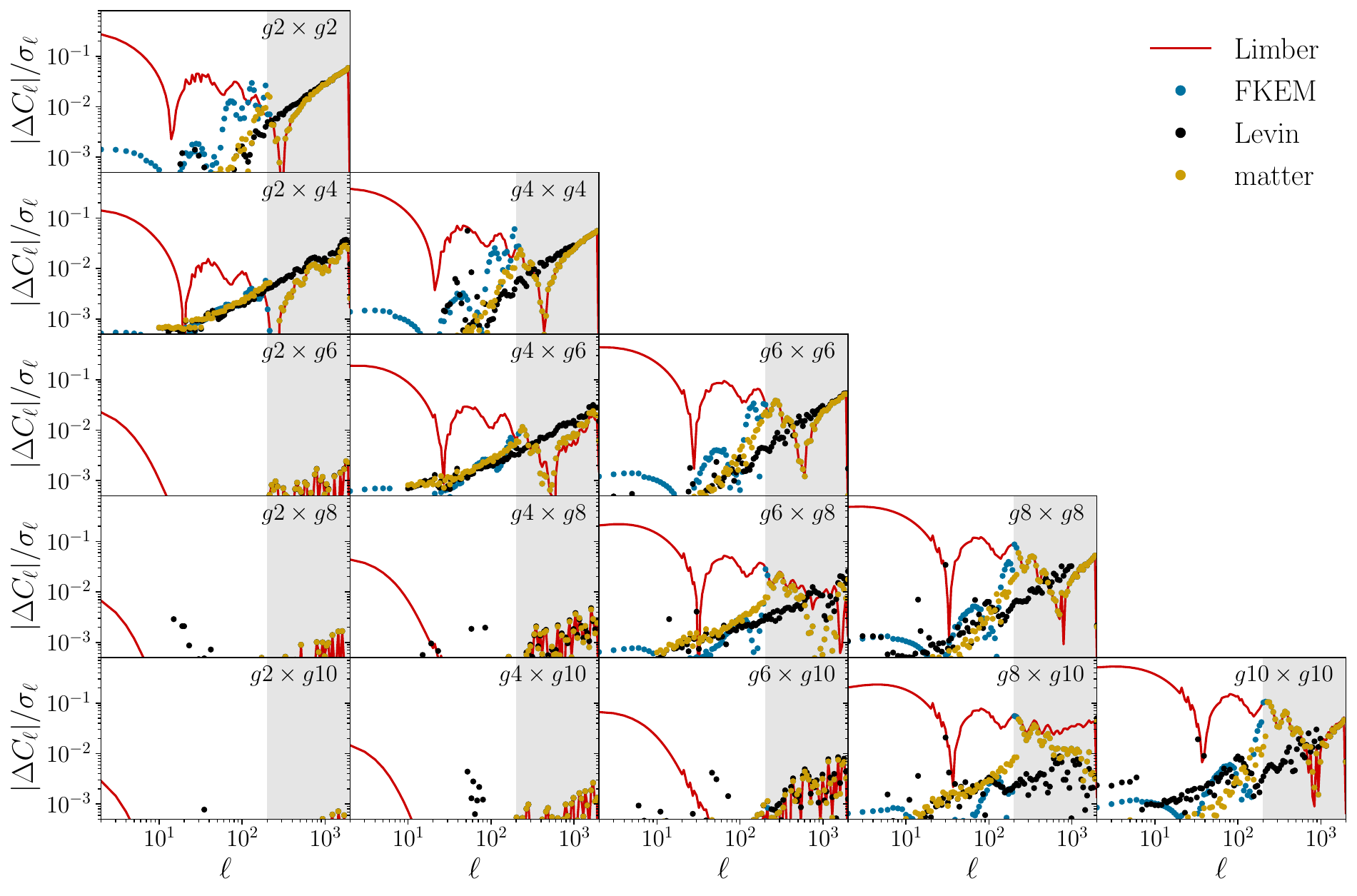}
  \caption{Deviation with respect to the benchmarks in the different auto- and cross-correlations between five of the clustering redshift bins used in this analysis as a fraction of the Gaussian uncertainties. Results are shown for the pure Limber approximation (red), and for the three non-Limber methods compared in the N5K challenge: \fkem (blue), \levin (black), and \matter (yellow). The grey band represents the region where $\ell>200$,  not part of the challenge set-up.}
\label{fig:dcl_gg}
\end{figure*}
\begin{figure*}
  \centering
  \includegraphics[width=0.85\textwidth]{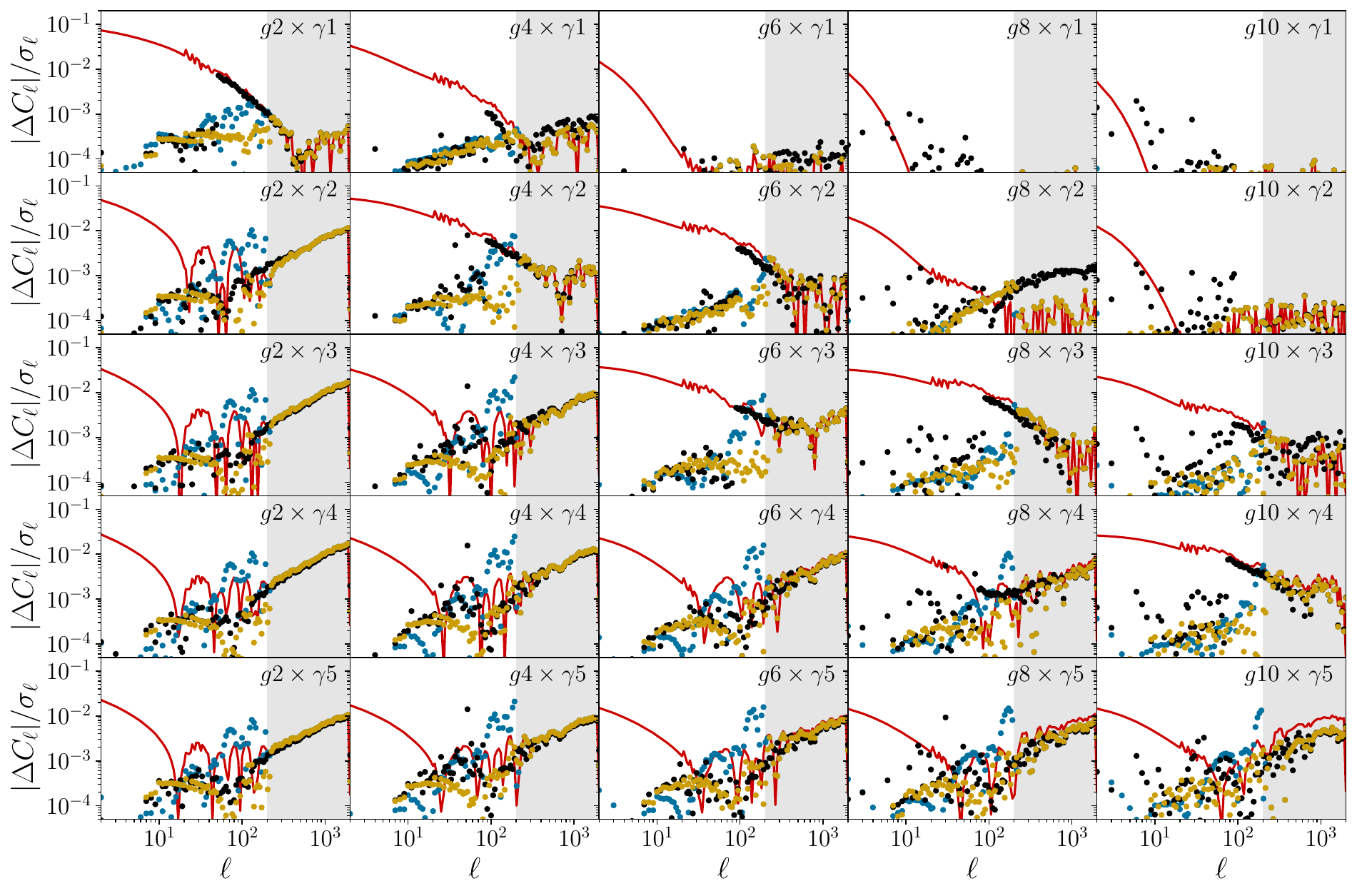}
  \caption{Same as Fig. \ref{fig:dcl_gg} for the clustering-shear cross-correlations (for the same clustering bins).}
\label{fig:dcl_gs}
\end{figure*}
\begin{figure*}
  \centering
  \includegraphics[width=0.85\textwidth]{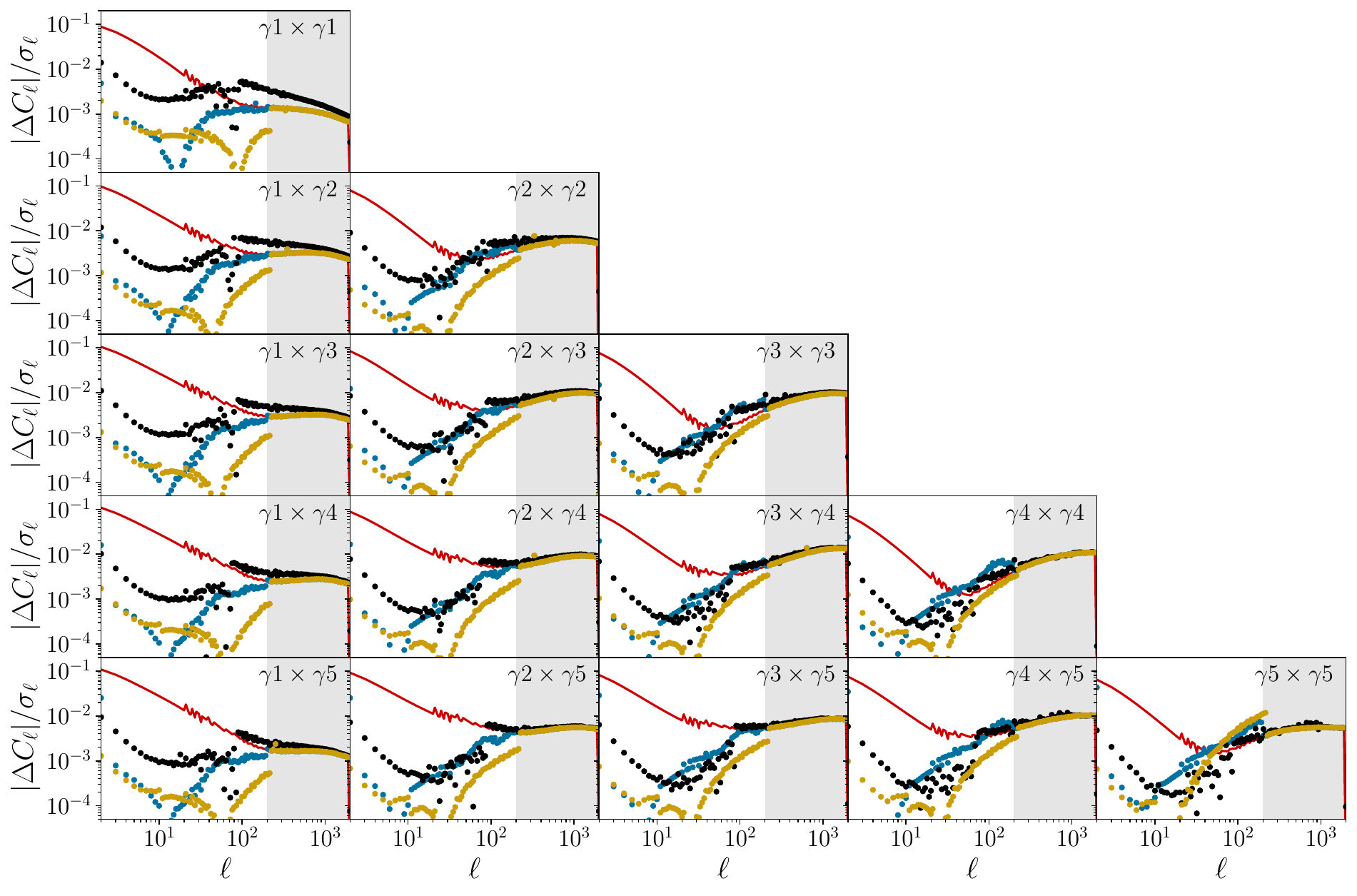}
  \caption{Same as Fig. \ref{fig:dcl_ss} for all the shear-shear auto- and cross-correlations.}
\label{fig:dcl_ss}
\end{figure*}
Our fiducial evaluation metric, as discussed in Section~\ref{subsec:metrics}, is the time to compute the LSST Y10 3x2pt data vector in the analysis scenario described in Section~\ref{subsec:config} to an accuracy of $\Delta \chi^2 \approx 0.2$ on $\ell \le 200$, using 64 cores on a single NERSC Haswell node. The fiducial run-time results are (with the run-time for the current implementation in CCL using the Limber approximation, for comparison): 
\begin{itemize}
\item{{\tt FKEM}: t = (0.437 $\pm$ 0.0005) s}
\item{{\tt Levin:} t=  (4.53 $\pm$ 0.007) s}
\item{{\tt matter:} t = (2.00 $\pm$ 0.009) s}
\item{{\tt CCL} (Limber): t = (0.0445 $\pm$ 0.0004) s}
\end{itemize}

Unless otherwise stated, mean run time values and their uncertainties here and throughout are computed via a set of 10 run times for each set-up. The uncertainty quoted is on the mean and as such is given by $\sigma = \frac{\sigma_{\rm std}}{\sqrt{10}}$ where $\sigma_{\rm std}$ is the standard deviation of the 10 runs. Note that entrants had the option of separating out cosmology-independent calculations which would need only to be performed once at the start of an MCMC-type analysis; the run-times presented here exclude those calculations labeled as separable in this way.

We clearly see that for the fiducial analysis scenario and required accuracy level, \fkem is the fastest non-Limber algorithm. Its run time is well within the regime of what is tractable for a single theory vector evaluation within an MCMC type analysis (being a few seconds or less). Although in the fiducial setup \levin and \matter take considerably longer than {\tt FKEM},  their run times do not fall outside this rough guideline of feasibility for their use in MCMC. Thus, we now investigate how the three methods compare along other axes.

Before we move on to explore the performance of these algorithms in different scenarios, it is worth studying their differences with respect to the benchmarks at the level of the individual power spectra. Fig. \ref{fig:dcl_gg} shows the difference with respect to the benchmarks as a fraction of the Gaussian statistical uncertainties, for the auto- and cross-correlations between 5 of the clustering redshift bins. Results are shown for the Limber approximation (red) and for the three non-Limber methods explored here (blue, black and yellow for \fkem, \levin, and \matter respectively). The $\Delta\chi^2$ of the Limber prediction itself with respect to the benchmark calculation is dominated by differences in the auto-correlations in $z$-bins as well as in the cross-correlation between adjacent bins; this discrepancy is observed at low $\ell$ to be at the level of a few ten per-cent. All non-Limber methods compared here are able to reduce these low-$\ell$ differences to irrelevant sub-percent levels on these scales. Interestingly, on intermediate scales ($\ell\sim100$), some of the non-Limber methods achieve a lower accuracy than the Limber approximation, although this does not affect the final $\Delta\chi^2$ significantly. At high $\ell$, all codes achieve accuracies comparable to the Limber approximation (unsurprisingly, since some of the methods indeed resort to this approximation at high $\ell$). Figs. \ref{fig:dcl_gs} and \ref{fig:dcl_ss} show the same results for the clustering-shear and shear-shear correlations respectively. Qualitatively, the conclusions are similar. However, it must be noted that, as could be expected given the width of the weak-lensing kernel, the relative differences with the Limber approximation are consistently smaller (a few $\%$ of the uncertainties at most) than in the case of the clustering correlations, and they increase less sharply at high $\ell$.

\subsection{Deviations from the fiducial setup}
\label{subsec:scaling}
\subsubsection{Required accuracy}
\label{subsubsec:varydchi2}
\noindent
We first consider how the evaluation time of each entry scales with the maximum allowed $\Delta \chi^2$ at $\ell \le 200$. As mentioned above, the fiducial required accuracy has been selected because it guarantees that any inaccuracy due to non-Limber integration will not induce a spurious $1\sigma$ (or greater) detection of a new effect characterised by a linear theory model. This is a conservative choice as it is robust against even the worst-case scenario, where the impact of non-Limber integration on the signal closely mimics that of the effect at risk of being spuriously detected. In reality, we are likely able to tolerate slightly less stringent accuracy requirements (a conjecture which can be verified for a given set-up with simulated parameter inference analyses). Given this possibility, we are interested to see how the run time of the different methods under consideration scales with required accuracy within the regime of accuracy which may be tolerable. We examine the behaviour of each method as we loosen the requirement on $\Delta \chi^2$ from the fiducial case to a most relaxed case of $\Delta \chi^2 \le 1.7$ on $\ell \leq 200$. This range of max allowed $\Delta \chi^2$ values is chosen because it corresponds to allowing for a spurious detection of a new effect characterised by a linear-theory model of roughly $1.5\sigma$ in the worst-case scenario. 

\begin{figure*}
\centering
\begin{minipage}[b]{.45\linewidth}
\includegraphics[width=\textwidth]{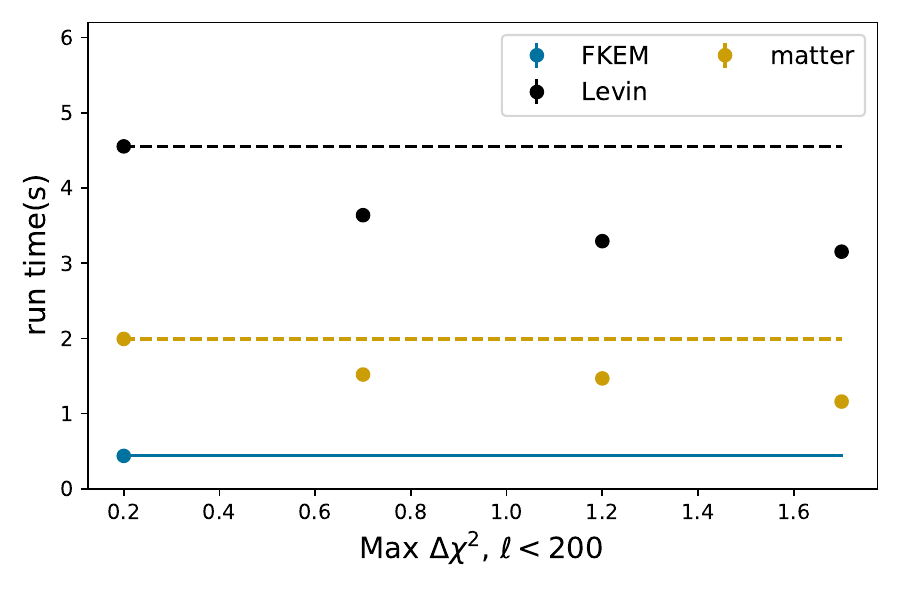}
\caption{Run-time on 64 cores as a function of maximum allowed $\Delta \chi^2$ on $\ell \le 200$. The {\tt FKEM} method does not present significant run time differences on this range of allowed $\Delta \chi^2$, so we present only the fiducial $\Delta \chi^2$ point and a horizontal line for visual comparison with other methods. The fiducial case for the other two cases is also displayed as dashed lines. Uncertainties (on the mean run times) are too small to be visible compared to the size of points.}
\label{fig:varydchi2}
\end{minipage}\hspace{0.5cm}
\begin{minipage}[b]{.45\textwidth}
\includegraphics[width=\textwidth]{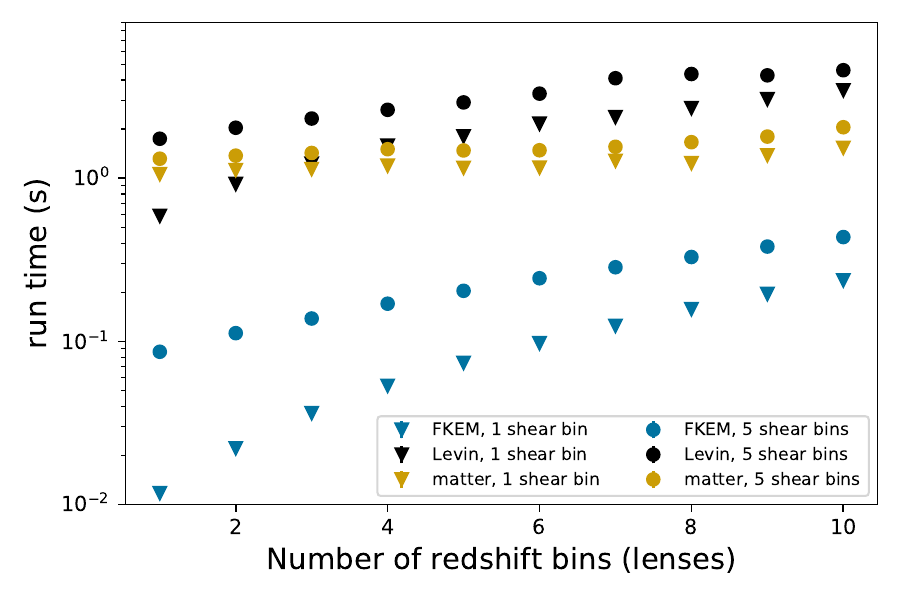}
\caption{Run time as a function of the number of clustering bins (and hence the number of spectra computed) for fixed number of shear bins.}
\label{fig:numbins_fixshear}
\end{minipage}
\end{figure*}

The resulting run times are displayed in Figure \ref{fig:varydchi2}. Note that it was determined that the {\tt FKEM} method did not benefit meaningfully in run time from changes to precision settings which would have corresponded to a loosening of $\Delta \chi^2$ requirements, so the $\Delta \chi^2 \le 0.2$ result only is displayed in that case as a horizontal line. We see that as expected, {\tt Levin} and {\tt matter} run times do decrease with the relaxation of required accuracy. To be specific, this relaxation is achieved for \matter by decreasing the number of FFT components ($N$ in equation \ref{eq:pl_mat}); this is the only option available which has an impact on timing in the \matter case as all other precision parameters were already set to their most relaxed acceptable value in the fiducial evaluation and overhead-dominated at these values. In the case of \levin, relaxation of the accuracy requirements was achieved mainly by allowing the extended Limber approximation to be used at lower $\ell$. Reducing the number of collocation points from 10 to 9 and relaxing error tolerances impacted timing somewhat but to a lesser extent than reducing the number of $\ell$-modes for which the full non-Limber calculation was being used. The resulting reduction in run-time is significant in the context of MCMC speed requirements: {\tt Levin} run time is reduced to less than 4s, and the {\tt matter} run time approaches 1s vs 2s in the fiducial case. However, in neither case are the fiducial sub-half-second times of {\tt FKEM} approached, meaning that {\tt FKEM} remains the preferred entry under this test.

\subsubsection{Number of spectra}
\noindent
We now examine the possibility that the run time of the different methods of non-Limber integration scale differently with the number of auto- and cross-spectra to be computed. To do so, we fix the number of shear bins while varying the number of clustering bins, which will deviate the most from the regime of Limber validity. We did perform the inverse check (fixing the number of clustering bins while varying shear bins) and found no qualitative differences in behaviour. 

When varying the number of redshift bins, we do not `re-bin' the overall redshift distribution. We work with a fixed set of 10 redshift bins for lens galaxies (clustering) and 5 for source galaxies (shear), as defined in Section \ref{subsec:config}. When considering fewer than all of these redshift bins, we simply include only the $n$ highest-redshift bins of this fiducial case. For example, when considering the case with $6$ lens bins, these are the 6 highest-redshift bins from the fiducial set of 10, neglecting the 4 lowest-redshift bins.  We examine both the case where all 5 fiducial shear bins are included, as well as the case where we include just the single highest-redshift shear bin. For this subsection only, run time results are averaged over 100 runs (rather than 10), as the small number of bins introduced more instability in the timing results.

The results are shown in Figure \ref{fig:numbins_fixshear}. We see that {\tt FKEM} and {\tt Levin} both present significant scaling of run time vs number of clustering bins. {\tt matter}, on the other hand is relatively flat. A possible explanation of this is that for \matter we are in the regime where the run-time is dominated by `overhead' calculations which must be carried out regardless of the number of spectra $N$ to be computed. If this is the case, we would expect that we would enter a regime where \fkem becomes slower than \matter, as \fkem scales strongly with $N$. Going to very high $N$ would likely then ultimately see the opposite behaviour, as \matter is expected to scale like $N^2$ (once beyond the regime of being dominated by $N$-independent overhead), whereas the factorisability of \fkem allows it to scale like $N$. Verifying this conjectured explanation would require exercising the \fkem and \matter methods in further scenarios with larger numbers of auto- and cross-spectra. As this is not the relevant scenario to an LSST 3x2pt analysis, we deem this to be beyond the scope of this paper but consider it to be of potential interest for future work, particularly in the context of spectroscopic surveys where more tomographic bins would be feasible.  

\subsubsection{Width of redshift bins}
\noindent
Employing tomographic bins of galaxies which are narrower in redshift space, particularly for the clustering sample, will naturally lead our analysis to fall further from the regime of validity of the Limber approximation, as discussed in Section~\ref{sec:theory} above. This type of deviation from the fiducial analysis may interact differently with the three methodologies under consideration, so we now consider this possibility. 

We construct two new analysis set-ups which differ from the fiducial scenario by, essentially, reducing the redshift width of each bin by a factor of two, and by a factor of four. Because of the joint impact of modelling photo-z scatter as well as imposing a redshift-selection function, this in effect takes the following form:
\begin{itemize}
    \item{{\bf Half-width:} We reduce the $z$-extent of each top-hat selection function by a factor of two and reduce both shear and clustering photo-z uncertainty by a factor of 5 (photo-z uncertainty is given by $\sigma_z(1+z)$, $\sigma_z$ is reduced by a factor of 5). The photo-z uncertainty is chosen such that the top-hat width dominates the overall width of the bin, such that in this sense the bin is half the width of the fiducial scenario.}
    \item{{\bf Quarter-width:} Identical to half-width except the top-hat selection function is reduced in $z$-extent by a factor of 4 ($\sigma_z$ is the same as in the half-width case).}
\end{itemize}

In this case, we are most interested in how the accuracy of the method may degrade as we push further into the non-Limber regime (rather than the run time). In initial tests, we found that the precision parameters set in the fiducial case for {\tt FKEM} failed dramatically in this scenario, leading to behaviour in which the $\Delta \chi^2$ peaked for the half-width case as compared to the full-width or quarter-width and indicating a lack of convergence of the method with respect to these parameters. We therefore altered the precision parameters of {\tt FKEM} for this test, doubling the number of points at which the $\chi$ integral is sampled (`{\tt Nchi\_fft}') and the number of zeros used to extend the high and low ends of the integrand (`{\tt Npad\_fft}'). In the fiducial setup, this change in precision parameters corresponded to a run time of 0.65\,s, still comfortably below the other two entries.

\begin{figure*}
\centering
\begin{minipage}[b]{.45\linewidth}
\includegraphics[width=\textwidth]{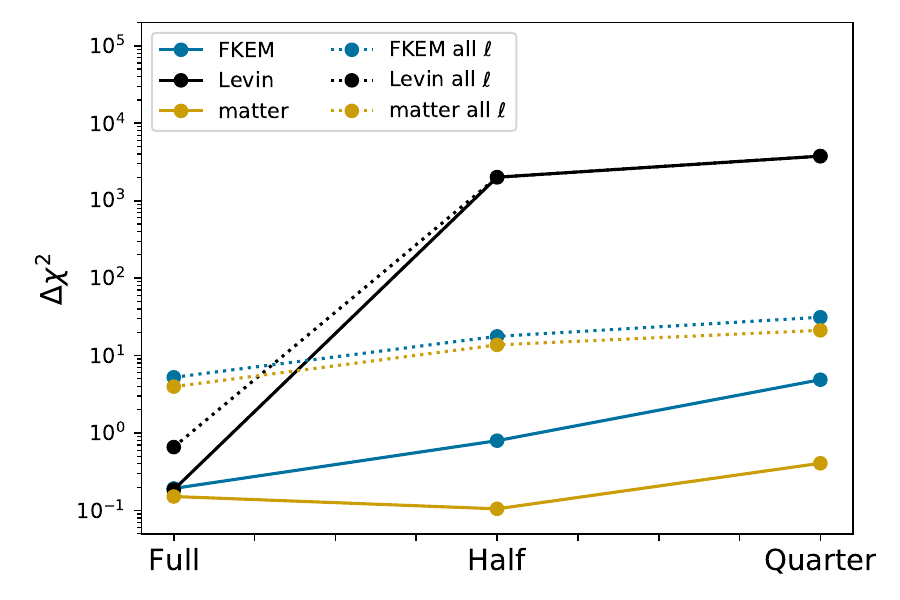}
\caption{The impact on the accuracy (as measured by $\sqrt{\Delta \chi^2}$ for $\ell<200$ and $\ell<2000$) achieved by each integration method as a function of the width of the clustering sample bins relative to the fiducial case.}
\label{fig:width}
\end{minipage}\hspace{0.5cm}
\begin{minipage}[b]{.45\textwidth}
\includegraphics[width=\textwidth]{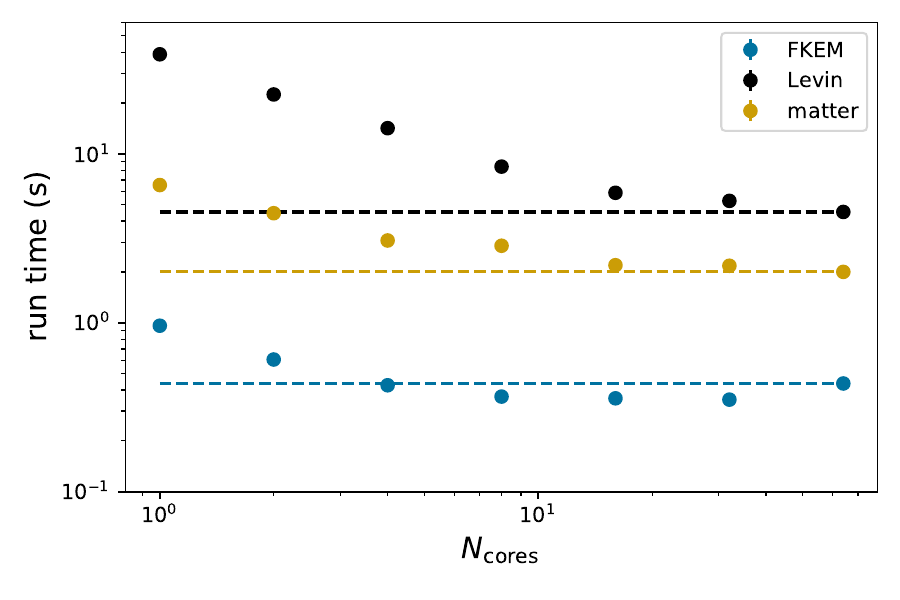}
\caption{The impact on run time as a function of the number of threads available on a single node. Fiducial results for 64 cores are shown as dashed horizontal lines.}
\label{fig:cores}
\end{minipage}
\end{figure*}

The impact of bin-width on the accuracy of the three methods is shown in Figure \ref{fig:width}. Here, we plot $\sqrt{\Delta \chi^2}$ -- essentially, the spurious signal-to-noise as compared to the benchmark data vector. As expected, the reduction of the bin widths generally results in a decrease of the accuracy of all the methods, however the degree of impact differs. {\tt matter} experiences near-negligible increased contribution to the $\sqrt{\Delta \chi^2}$ from $\ell\le 200$ when moving to narrower bins, while the accuracy of the {\tt Levin} method is reduced dramatically in moving from the fiducial scenario to half-width. {\tt Levin} also displays the peculiar behaviour that in the half- and quarter-width case, the accuracy degradation seems to be near-exclusively sourced in the $\ell \le 200$ regime. Further exploration of this behaviour reveals that this is due to the fact that in {\tt Levin}'s default configuration, the full non-Limber calculation (in operation on this lower $\ell$ range) fails dramatically for these narrower bins. Above $\ell=200$, however, {\tt Levin} uses second-order Limber, which is relatively well-behaved even for the narrower bins considered here. One could consider modifying the precision parameters of {\tt Levin} for this test only, as for {\tt FKEM}, however given {\tt Levin}'s already-higher runtime, this is unlikely to produce a configuration which remains viable for, for example, an MCMC analysis. Finally, {\tt FKEM} fares worse than {\tt matter} for $\ell \le 200$ alone, but the two methods perform comparably across the full $\ell$ range. 

\subsubsection{Number of cores}
\noindent
In a computationally heavy task such as high-dimensional parameter inference, we often want to take advantage of parallelisation to speed up calculations. As a proxy for the parallelisability of each entry, we consider how run-time scales with number of cores available. Our fiducial analysis setup uses a single node with 32 physical cores, each double-threaded. We refer to the full 32x2 node as 64 threads. 

The variation in run time of each entry with number of cores available (all on a single node) is shown in Figure \ref{fig:cores}. As expected, all three methods experience a reduction in run time with an increase in available cores, with the effect on {\tt Levin} being arguably the most significant. {\tt FKEM} in fact experiences a slight increase in run time as we move to the maximum number of cores considered; we interpret this as being the result of the fact that setting up more cores does require some overhead, and since {\tt FKEM} as submitted is not extensively parallelised, this overhead is actually the limiting factor at higher core numbers. We note that more than the other aspects of the evaluation, the question of scaling with number of cores depends on implementation details. It is likely that each entry could improve its performance in this respect without significantly altering the fundamental algorithm. 

\section{Conclusions}
\label{sec:conc}

\noindent
A rapid and accurate implementation of non-Limber integration for the calculation of the 3x2pt data vector is an essential component of the cosmological inference pipeline for LSST. In this work, we have examined three methods and corresponding implementations to achieve this goal: {\tt Levin}, a linear-algebra based method, {\tt matter}, which uses FFT in logspace to perform the required integral over the double Bessel Functions, and {\tt FKEM}, which first splits the matter power spectrum into additive linear and nonlinear components, and then uses this split to cast the required integral as over individual Bessel functions, which are then also solved via FFTlog.

Our investigation finds that in terms of speed of computation at a required accuracy which is in the regime of what we require for LSST, the {\tt FKEM (CosmoLike)} method is the preferred method of the three. We find that {\tt FKEM} maintains its run time advantage while varying the number of redshift bins included in the analysis (and hence the number of auto- and cross-spectra to calculate), as well as when considering reasonable relaxations to our fiducial accuracy requirement. 

There are however certain metrics by which one could argue for {\tt Levin} or {\tt matter} as being superior: while {\tt FKEM} maintains its speed advantage for the range of number of spectra considered, {\tt matter}'s run time appears to be less sensitive to increasing this number, potentially advantaging it in high bin number analyses. Of the three entries, {\tt FKEM} also does not have the most efficient parallelisation, and in fact seems to lose time when running on more cores above a certain moderate fraction of a node. {\tt FKEM} and {\tt matter} also perform comparably in accuracy when moving to narrower redshift bins, with {\tt matter} maintaining a better accuracy on $\ell\le 200$.

Although {\tt FKEM} does come out on top in many ways in this challenge, one scenario in which it presents a challenge is the case in which the factorisation of $P_\delta$ into the form of equation \ref{eq:fkem_factor} is not straightforward. This could arise within cosmological models beyond $w$CDM, such as in alternative theories of gravity with scale-dependent growth. In a similar vein, the formulation of the {\tt FKEM} method presented here would require modification to account for primordial non-Gaussianity introducing scale-dependence on linear scales. It is most likely possible to circumvent this issue by considering narrow redshift bins in which these effects are approximately scale-independent for a given bin, as discussed in \citep{Fang2020}. However, this would require some non-trivial development and the impact on the accuracy of calculations would need to be studied across all relevant scenarios. The fact that neither {\tt matter} nor {\tt Levin} rely on this factorisation is thus a potential advantage in these very realistic analysis scenarios. 

Of course, there are also other more general advantages of the availability of multiple reliable and well-tested non-Limber integration codes. The development of multiple independent codes for the same theoretical calculations has long been a driving force in cosmological analyses, allowing for crucial validation of complex calculation (see e.g. Boltzmann codes {\tt CLASS} \citep{class} and {\tt CAMB} \citep{camb}). We therefore anticipate a potential integration of more than one of these methods into the LSST DESC pipeline software {\tt CCL}. Another possible avenue of future work which we have not explored here is emulating the calculation of the non-Limber integrals. While we do not deem emulation to at present have significant advantages over the numerical integration methods described here, this may be of interest in the future as the development of fast emulation techniques continues to accelerate \citep[see e.g.][for just a few examples]{Euclid2019, Ramachandra2021}. 

In this work, we focus on 3x2pt, as this will constitute a key analysis for LSST DESC static science. However, given that the Limber approximation breaks down more readily for galaxy clustering (with its narrower kernel), one might wonder whether for cosmic-shear-only analyses, the Limber approximation might suffice for the LSST Y10 scenario considered here. We find that it does not: the Limber approximation cannot achieve the accuracy requirement of the challenge when considering only the cosmic shear portion of the data vector, in qualitative agreement with the findings of \cite{Kilbinger2017} on this point for a different Stage IV survey (Euclid).

One unexpected finding of the evaluation of this challenge was the fact that using the standard (first-order) Limber approximation on $200 \le \ell \le 1000$ resulted in significant values of $\Delta \chi^2$ for the fiducial 3x2pt Y10 LSST analysis. This was contrary to our expectations based on conventional wisdom in the field that the first-order Limber approximation would certainly be adequate above $\ell=200$. As a result, an implementation of the second-order Limber approximation, which does reduce $\Delta \chi^2$ to an acceptable level on $200 \le \ell \le 1000$ for LSST Y10, is planned to be implemented in {\tt CCL} alongside full non-Limber support.

\begin{acknowledgments}
This paper has undergone internal review in the LSST Dark Energy Science Collaboration. The internal reviewers were Elisa Chisari, David Kirkby, and Joe Zuntz; we thank them for their helpful comments and discussion.

Author contributions: CDL led the challenge development and organisation, led the evaluation of challenge entries and other analysis, and wrote much of the text for this paper. TF contributed significantly to the evaluation and analysis of challenge entries. XF, RR, NS, and TT were the authors of the three challenge entries, assisted with the iterative evaluation and analysis process, and wrote text for this paper especially Section 4. DA wrote code required for the challenge infrastructure, produced the benchmarks, contributed to the evaluation and analysis, and wrote some text for this paper. JEC, FL, and AS contributed to the initial design of the challenge, discussions, and made smaller contributions to the evaluation and analysis (e.g. JEC used \texttt{Angpow} to cross-check some benchmark outputs). MI is a DESC Builder and made significant contributions to the development of CCL on which the code infrastructure for the challenge relies. 

XF acknowledges the support of fellowship by the Berkeley Center for Cosmological Physics.
NS acknowledges the support of the following Maria de Maetzu fellowship grant: Esta publicaci\'on es parte de la ayuda CEX2019-000918-M, financiada por MCIN/AEI/10.13039/501100011033.
RR is supported by the European Research Council (Grant No. 770935).
TT acknowledges support from the Leverhulme Trust.
TF acknowledges support from INCT e-Universo.
DA acknowledges support from the Science and Technology Facilities Council through an Ernest Rutherford Fellowship, grant reference ST/P004474. 

We made use of computational resources at the University of Oxford Department of Physics, funded by the John Fell Oxford University Press research fund.
The DESC acknowledges ongoing support from the Institut National de 
Physique Nucl\'eaire et de Physique des Particules in France; the 
Science \& Technology Facilities Council in the United Kingdom; and the
Department of Energy, the National Science Foundation, and the LSST 
Corporation in the United States.  DESC uses resources of the IN2P3 
Computing Center (CC-IN2P3--Lyon/Villeurbanne - France) funded by the 
Centre National de la Recherche Scientifique; the National Energy 
Research Scientific Computing Center, a DOE Office of Science User 
Facility supported by the Office of Science of the U.S.\ Department of
Energy under Contract No.\ DE-AC02-05CH11231; STFC DiRAC HPC Facilities, 
funded by UK BEIS National E-infrastructure capital grants; and the UK 
particle physics grid, supported by the GridPP Collaboration.  This 
work was performed in part under DOE Contract DE-AC02-76SF00515.
We acknowledge the use of the Python libraries {\tt NumPy} \citep{numpy, van2011numpy} and {\tt matplotlib} \citep{Hunter:2007} as well as the GNU Scientific Library {\tt GSL} in this work.
\end{acknowledgments}

\bibliographystyle{apalike}

\providecommand{\noopsort}[1]{}\providecommand{\singleletter}[1]{#1}%

\end{document}